\documentclass[12pt]{article}
\usepackage{amssymb,amsfonts}
\usepackage{amsmath}
\usepackage{epsf,epsfig}
\begin{document}
\baselineskip 18pt

\title{$R$-matrix for the XX spin chain at the special imaginary value of staggered magnetic field}
\author{P.~N.~Bibikov}
\date{\it Russian State Hydrometeorological University, Saint-Petersburg, Russia}

\maketitle

\vskip5mm

\begin{abstract}
It is pointed, that the $16\times16$ Hamiltonian density matrix, corresponding to XX spin chain in a staggered magnetic field, satisfies
the Reshetikhin condition as well, as the two higher ones. All of them are necessary for the existence of the corresponding $R$-matrix.
At the special imaginary value of the staggered field, the $R$-matrix is obtained explicitly. It is shown that the corresponding Hamiltonian has purely imaginary spectrum.
\end{abstract}

\maketitle
\vskip20mm

The XX model, suggested in \cite{1}, and related to the Hamiltonian
\begin{equation}\label{hamgen}
\hat H=\sum_nH_{n,n+1},
\end{equation}
where
\begin{equation}\label{loc0}
H_{n,n+1}=\frac{1}{2}\Big({\bf S}^+_n{\bf S}^-_{n+1}+{\bf S}^-_n{\bf S}^+_{n+1}\Big),
\end{equation}
is now one of the most studied integrable spin chains (see \cite{2} and references therein). Here ${\bf S}^j_n$ ($j=+,-,z$) are the spin-1/2 operators
associated with the $n$-th site of the chain.

The integrability of \eqref{hamgen}, \eqref{loc0} has been first demonstrated within the Jordan-Wigner transformation to the free-fermion model and then
confirmed by various alternative approaches (see \cite{2} and references therein). Namely, the Hamiltonian density $4\times4$ matrix
\begin{equation}\label{dens0}
H=\frac{1}{2}\Big({\bf S}^+\otimes{\bf S}^-+{\bf S}^-\otimes{\bf S}^+\Big),
\end{equation}
related to \eqref{loc0}, may be represented in the form
\begin{equation}\label{hr}
H=\frac{1}{2}\frac{dR(\lambda)}{d\lambda}\Big|_{\lambda=0},
\end{equation}
where the $R$-matrix
\begin{equation}
R(\lambda)=\left(\begin{array}{cccc}
\cos{\lambda}&0&0&0\\
0&1&\sin{\lambda}&0\\
0&\sin{\lambda}&1&0\\
0&0&0&\cos{\lambda}
\end{array}\right),
\end{equation}
satisfies the Yang-Baxter equation (in the Braid group form) \cite{4}
\begin{equation}\label{YB}
R_{12}(\lambda-\mu)R_{23}(\lambda)R_{12}(\mu)=R_{23}(\mu)R_{12}(\lambda)R_{23}(\lambda-\mu),
\end{equation}
(as usual, $R_{12}(\lambda)\equiv R(\lambda)\otimes I_2$ and $R_{23}(\lambda)\equiv I_2\otimes R(\lambda)$) and the initial condition
\begin{equation}\label{init}
R(0)=I_4.
\end{equation}
Here by $I_M$ we denote the $M\times M$ identity matrix.

Within the slightly modified version of the \cite{1} approach, it has been proved \cite{3}, that the model \eqref{hamgen}, \eqref{loc0} remains integrable even under addition of a staggered longitudinal magnetic field, when
\begin{equation}\label{loc}
H_{n,n+1}(h_{\rm st})=\frac{1}{2}\Big[{\bf S}^+_n{\bf S}^-_{n+1}+{\bf S}^-_n{\bf S}^+_{n+1}+h_{\rm st}\Big((-1)^n{\bf S}^z_n+(-1)^{n+1}{\bf S}^z_{n+1}\Big)\Big],
\end{equation}
instead of \eqref{loc0}. Namely, with the use of the Fermi operators representation
\begin{equation}
{\bf S}^+_n=\prod_{m<n}\sigma_ma_n,\qquad{\bf S}^-_n=\prod_{m<n}\sigma_ma_n^{\dagger},\qquad{\bf S}^z_n=\frac{1}{2}[a_n,a_n^{\dagger}]
=\frac{1}{2}-a_n^{\dagger}a_n,
\end{equation}
where $\sigma_n\equiv1-2a_n^{\dagger}a_n$ and $\{a_m,a_n\}=0$, and $\{a_m,a_n^{\dagger}\}=\delta_{mn}$, the Hamiltonian \eqref{hamgen}, \eqref{loc0}
takes the well tractable bilinear form
\begin{equation}\label{hamf1}
\hat H(h_{\rm st})=\frac{1}{2}\sum_n\Big[a_{n+1}^{\dagger}a_n+a_n^{\dagger}a_{n+1}-h_{\rm st}
\Big((-1)^na_n^{\dagger}a_n+(-1)^{n+1}a_{n+1}^{\dagger}a_{n+1}\Big)\Big],
\end{equation}
which, under the substitutions $a_{2n}\equiv b_n$, and $a_{2n+1}\equiv\tilde b_n$, turns into
\begin{equation}\label{hamf2}
\hat H(h_{\rm st})=\frac{1}{2}\sum_n\Big(\tilde b_n^{\dagger}b_n+b_n^{\dagger}\tilde b_n+b_{n+1}^{\dagger}\tilde b_n+\tilde b_n^{\dagger}b_{n+1}
+h_{\rm st}
(2\tilde b_n^{\dagger}\tilde b_n-b_n^{\dagger}b_n-b_{n+1}^{\dagger}b_{n+1})\Big).
\end{equation}

With the use of the Fourier representations
\begin{equation}
b_n=\frac{1}{\sqrt{2\pi}}\int_0^{2\pi}dk{\rm e}^{-ikn}c_1(k),\qquad \tilde b_n=\frac{1}{\sqrt{2\pi}}\int_0^{2\pi}dk{\rm e}^{-ikn}c_2(k),
\end{equation}
where
\begin{equation}\label{comc12}
\{c_j^{\dagger}(k),c_l(p)\}=\delta_{jl}\delta(k-p),\qquad\{c_j(k),c_l(p)\}=0,
\end{equation}
\eqref{hamf1} is reduced to the form
\begin{equation}\label{Hc}
\hat H(h_{\rm st})=\int_0^{2\pi}dk
\left(\begin{array}{cc}
c_1^{\dagger}(k)&c_2^{\dagger}(k)
\end{array}\right)
M(k,h_{\rm st})
\left(\begin{array}{c}
c_1(k)\\
c_2(k)
\end{array}\right),
\end{equation}
where
\begin{equation}
M(k,h_{\rm st})=\frac{1}{2}\left(\begin{array}{cc}
-2h_{\rm st}&1+{\rm e}^{ik}\\
1+{\rm e}^{-ik}&2h_{\rm st}
\end{array}\right).
\end{equation}
As it has been shown in \cite{3}, for real $h_{\rm st}$ the Hamiltonian \eqref{Hc} may be diagonalized by unitary transformation, so that the model \eqref{hamgen}, \eqref{loc0} is integrable.

It is desirable to confirm this result by presentation of the corresponding $R$-matrix.
Since \eqref{loc} is invariant only under even translations ($H_{n,n+1}=H_{n+2,n+3}\neq H_{n+1,n+2}$), the approach based on a $4\times4$ matrix needs a modification.
Namely, one should pass from \eqref{dens0} to the $16\times16$ matrix, corresponding to the representation
\begin{equation}\label{hamgenm}
\hat H(h_{\rm st})=\sum_n{\cal H}_{2n,2n+3}(h_{\rm st}),
\end{equation}
more general than \eqref{hamgen}. Here
\begin{equation}\label{calH=}
{\cal H}_{2n,2n+3}(h_{\rm st})=\frac{1}{2}\Big(H_{2n,2n+1}(h_{\rm st})+H_{2n+2,2n+3}(h_{\rm st})\Big)+H_{2n+1,2n+2}(h_{\rm st}),
\end{equation}
and the corresponding $16\times16$ Hamiltonian density matrix is
\begin{equation}\label{Hst}
{\cal H}(h_{\rm st})=\frac{1}{2}\Big(H^{(+)}(h_{\rm st})\otimes I_4+I_4\otimes H^{(+)}(h_{\rm st})\Big)+
I_2\otimes H^{(-)}(h_{\rm st})\otimes I_2,
\end{equation}
where
\begin{equation}
H^{(\pm)}(h_{\rm st})=\frac{1}{2}\Big[{\bf S}^+\otimes{\bf S}^-+{\bf S}^-\otimes{\bf S}^+\pm h_{\rm st}
\Big({\bf S}^z\otimes I_2-I_2\otimes{\bf S}^z\Big)\Big].
\end{equation}

The direct calculation shows, that the matrix \eqref{Hst} satisfies the Reshetikhin condition \cite{4}
\begin{equation}\label{Resh}
[{\cal H}(h_{\rm st})\otimes I_4+I_4\otimes{\cal H}(h_{\rm st}),[{\cal H}(h_{\rm st})\otimes I_4,I_4\otimes{\cal H}(h_{\rm st})]]=I_4\otimes{\cal K}-{\cal K}\otimes I_4,
\end{equation}
(where the $16\times16$ matrix ${\cal K}$ may be readily calculated).

According to \eqref{Resh}, the sum
\begin{equation}\label{Qdef}
\hat Q=\sum_n[{\cal H}_{2n-2,2n+1}(h_{\rm st}),{\cal H}_{2n,2n+3}(h_{\rm st})],
\end{equation}
is the first integral of motion
\begin{equation}
[\hat H,\hat Q]=0.
\end{equation}
In fact, $\hat Q$ does not depend on $h_{\rm st}$ and may be reduced to the form
\begin{equation}\label{Qfact}
\hat Q=\frac{1}{2}\sum_n[H_{n-1,n},H_{n,n+1}].
\end{equation}
Really, following \eqref{calH=},
\begin{equation}\label{comm}
[{\cal H}_{2n-2,2n+1}(h_{\rm st}),{\cal H}_{2n,2n+3}(h_{\rm st})]=\frac{1}{2}\Big([H_{2n-1,2n}^{(-)},H^{(+)}_{2n,2n+1}]
+[H^{(+)}_{2n,2n+1},H^{(-)}_{2n+1,2n+2}]\Big).
\end{equation}
At the same time,
\begin{eqnarray}\label{comms}
&&[H^{(+)}_{2n,2n+1},H^{(-)}_{2n+1,2n+2}]=[H_{2n,2n+1},H_{2n+1,2n+2}]+h_{\rm st}[{\bf S}^z_{2n+1},H_{2n,2n+1}-H_{2n+1,2n+2}],\nonumber\\
&&[H_{2n-1,2n}^{(-)},H_{2n,2n+1}^{(+)}]=[H_{2n-1,2n},H_{2n,2n+1}]+h_{\rm st}[{\bf S}^z_{2n},H_{2n,2n+1}-H_{2n-1,2n}].
\end{eqnarray}
Since $[{\bf S}^z_n+{\bf S}^z_{n+1},H_{n,n+1}]=0$, the substitution of \eqref{comms} into \eqref{comm} gives
\begin{eqnarray}\label{comm2}
&&[{\cal H}_{2n-2,2n+1}(h_{\rm st}),{\cal H}_{2n,2n+3}(h_{\rm st})]=\frac{1}{2}\Big([H_{2n-1,2n},H_{2n,2n+1}]
+[H_{2n,2n+1},H_{2n+1,2n+2}]\Big)\nonumber\\
&&+\frac{h_{\rm st}}{4}\Big([{\bf S}^z_{2n-1}-{\bf S}^z_{2n},H_{2n-1,2n}]-[{\bf S}^z_{2n+1}-{\bf S}^z_{2n+2},H_{2n+1,2n+2}]\Big).
\end{eqnarray}
Under the substitution of \eqref{comm2} into \eqref{Qdef}, the terms proportional to $h_{\rm st}$ are canceled, and one gets \eqref{Qfact}.

It may be readily proved by direct calculations, that additionally to \eqref{Resh} the Hamiltonian density matrix satisfies two higher order integrability conditions, presented in \cite{5} and \cite{6}.
Both of them also are necessary (but not sufficient) for representation of ${\cal H}(h_{\rm st})$ in the form \eqref{hr}, but with the $16\times16$ $R$-matrix ${\cal R}(\lambda,h_{\rm st})$, which
satisfies as the Yang-Baxter equation \eqref{YB}, as the initial condition \eqref{init} (with $I_4$ replaced by $I_{16}$).

As it has been suggested in \cite{6}, and shown in \cite{7,8}, in such a situation one may try to guess the entries of $R$-matrix by analyzing the series expansions of their ratios.
For the model \eqref{hamgenm} the author succeed, however, only in the two special cases, $h_{\rm st}=0$ and $h_{\rm st}=i$.

The direct analysis of power series for the entries shows, that both the $R$-matrices have the similar structure
\begin{equation}\label{template}
{\cal R}=\left(\begin{array}{cccccccccccccccc}
*&0&0&0&0&0&0&0&0&0&0&0&0&0&0&0\\
0&*&*&0&*&0&0&0&0&0&0&0&0&0&0&0\\
0&*&*&0&*&0&0&0&*&0&0&0&0&0&0&0\\
0&0&0&*&0&*&*&0&0&*&*&0&*&0&0&0\\
0&*&*&0&*&0&0&0&*&0&0&0&0&0&0&0\\
0&0&0&*&0&*&*&0&0&*&*&0&*&0&0&0\\
0&0&0&*&0&*&*&0&0&*&*&0&*&0&0&0\\
0&0&0&0&0&0&0&*&0&0&0&*&0&*&0&0\\
0&0&*&0&*&0&0&0&0&0&0&0&0&0&0&0\\
0&0&0&*&0&*&*&0&0&*&*&0&*&0&0&0\\
0&0&0&*&0&*&*&0&0&*&*&0&*&0&0&0\\
0&0&0&0&0&0&0&*&0&0&0&*&0&*&*&0\\
0&0&0&*&0&*&*&0&0&*&*&0&*&0&0&0\\
0&0&0&0&0&0&0&*&0&0&0&*&0&*&*&0\\
0&0&0&0&0&0&0&0&0&0&0&*&0&*&*&0\\
0&0&0&0&0&0&0&0&0&0&0&0&0&0&0&*\\
\end{array}\right),
\end{equation}
and for both of them
\begin{equation}\label{rels1}
{\cal R}_{ji}={\cal R}_{ij},\qquad{\cal R}_{17-i,17-j}=\bar{\cal R}_{ij},
\end{equation}
(for the shortness we put ${\cal R}_{ij}\equiv {\cal R}_{ij}(\lambda,h_{\rm st})$). Moreover,
\begin{eqnarray}\label{rels2}
&&{\cal R}_{16,16}={\cal R}_{11},\quad{\cal R}_{15,15}={\cal R}_{99},\quad {\cal R}_{14,14}={\cal R}_{55},\nonumber\\
&&{\cal R}_{12,12}={\cal R}_{33},\quad {\cal R}_{10,10}={\cal R}_{77},\quad {\cal R}_{8,8}={\cal R}_{22},\nonumber\\
&&{\cal R}_{14,15}={\cal R}_{59},\quad {\cal R}_{10,11}={\cal R}_{7,11},\quad {\cal R}_{10,13}={\cal R}_{7,13},\nonumber\\
&&{\cal R}_{8,12}={\cal R}_{23},\quad {\cal R}_{6,10}={\cal R}_{67},\quad {\cal R}_{4,10}={\cal R}_{47},\nonumber\\
&&{\cal R}_{7,10}={\cal R}_{39}={\cal R}_{8,14}={\cal R}_{12,15}={\cal R}_{25}
={\cal R}_{6,11}-{\cal R}_{4,13}={\cal R}_{77}-{\cal R}_{11},\nonumber\\
&&{\cal R}_{12,14}={\cal R}_{35}={\cal R}_{67}+{\cal R}_{10,11}.
\end{eqnarray}

According to \eqref{rels1}, it convenient to introduce the real variables
\begin{equation}\label{ABdef}
{\cal A}_{ij}=\frac{{\cal R}_{ij}+{\cal R}_{17-i,17-j}}{2},\qquad
{\cal B}_{ij}=\frac{{\cal R}_{ij}-{\cal R}_{17-i,17-j}}{2i},
\end{equation}
for which
\begin{equation}\label{ABrel}
{\cal A}_{ji}={\cal A}_{ij},\quad{\cal B}_{ji}={\cal B}_{ij},\quad{\cal A}_{17-i,17-j}={\cal A}_{ij},\quad{\cal B}_{17-i,17-j}=-{\cal B}_{ij}.
\end{equation}
Following \eqref{rels2},
\begin{eqnarray}\label{Arel}
&&{\cal A}_{55}={\cal A}_{33},\quad{\cal A}_{88}={\cal A}_{22},\quad A_{4,10}=A_{47},\quad{\cal A}_{59}={\cal A}_{23},\quad{\cal A}_{6,10}={\cal A}_{67},\nonumber\\
&&{\cal A}_{39}={\cal A}_{25}={\cal A}_{7,10}={\cal A}_{77}-{\cal A}_{11},\quad
{\cal A}_{4,13}={\cal A}_{6,11}+{\cal A}_{11}-{\cal A}_{77},\nonumber\\
&&{\cal A}_{7,11}={\cal A}_{35}-A_{67},
\end{eqnarray}
and
\begin{eqnarray}\label{Brel}
&&{\cal B}_{11}={\cal B}_{77}={\cal B}_{25}={\cal B}_{35}={\cal B}_{39}=0,\quad{\cal B}_{55}=-{\cal B}_{33},\quad{\cal B}_{88}={\cal B}_{22},\nonumber\\
&&{\cal B}_{59}=-{\cal B}_{23},\quad
{\cal B}_{4,10}={\cal B}_{47},\quad{\cal B}_{6,10}={\cal B}_{67},
\end{eqnarray}

According to \eqref{ABdef}-\eqref{Brel}, all the entries of the $R$-matrix ${\cal R}(\lambda,h_{\rm st})$ \eqref{template} are linear combinations of
13 parameters $\{{\cal A}_{11},{\cal A}_{22},{\cal A}_{33},{\cal A}_{44},{\cal A}_{66},{\cal A}_{77},{\cal A}_{23},{\cal A}_{35},{\cal A}_{46},{\cal A}_{47},{\cal A}_{4,11},{\cal A}_{67},{\cal A}_{6,11}\}$, and 9 parameters $\{{\cal B}_{22},{\cal B}_{33},{\cal B}_{44},{\cal B}_{66},{\cal B}_{23},{\cal B}_{46},{\cal B}_{47},{\cal B}_{4,11},{\cal B}_{67}\}$.
Namely, for $h_{\rm st}=0$ one may put
\begin{eqnarray}
&&{\cal A}_{11}={\tt c}^4,\quad{\cal A}_{22}={\tt c}^3,\quad{\cal A}_{33}={\tt c}(1+{\tt s}^2),\quad{\cal A}_{44}=1,\quad{\cal A}_{66}=1+{\tt s}^2-{\tt s}^4,\nonumber\\
&&{\cal A}_{77}={\tt c}^2,\quad{\cal A}_{23}={\tt c}^2{\tt s},\quad{\cal A}_{35}=2{\tt c}{\tt s},\quad{\cal A}_{46}=(2-{\tt s}^2){\tt s},\quad{\cal A}_{47}={\tt c}{\tt s}^2,\quad{\cal A}_{4,11}={\tt s}^3,\nonumber\\
&&{\cal A}_{67}={\tt c}{\tt s},\quad{\cal A}_{6,11}={\tt s}^2,\quad{\cal B}_{ij}=0,\qquad{\tt c}\equiv\cos{\lambda},\quad{\tt s}\equiv\sin{\lambda}.
\end{eqnarray}
At the same time, for $h_{\rm st}=i$
\begin{eqnarray}
&&{\cal A}_{11}={\cal A}_{22}={\tt ch}^4,\quad{\cal A}_{33}=(2-{\tt ch}^2){\tt ch}^2,\quad{\cal A}_{44}=2{\tt ch}^2-1,
\quad{\cal A}_{66}=1-3{\tt ch}^2{\tt sh}^2,\nonumber\\
&&{\cal A}_{77}=(2{\tt ch}^2-1){\tt ch}^2,\quad{\cal A}_{23}={\tt ch}^3{\tt sh},\quad{\cal A}_{35}=2{\tt ch}{\tt sh},\quad{\cal A}_{46}=(3-{\tt ch}^2){\tt ch}{\tt sh},\nonumber\\
&&{\cal A}_{47}=
{\tt ch}^2{\tt sh}^2,\quad{\cal A}_{4,11}={\tt ch}{\tt sh}^3,\quad{\cal A}_{67}={\tt ch}{\tt sh},\quad{\cal A}_{6,11}=(2{\tt ch}^2-1){\tt sh}^2,\nonumber\\
&&{\cal B}_{22}={\tt ch}^3{\tt sh},\quad{\cal B}_{33}=-(2+{\tt ch}^2){\tt ch}{\tt sh},\quad{\cal B}_{44}=-2{\tt ch}{\tt sh},\quad
{\cal B}_{66}=4{\tt ch}{\tt sh},\nonumber\\
&&{\cal B}_{23}=-{\tt ch}^2{\tt sh}^2,\quad{\cal B}_{46}=(1+{\tt ch}^2){\tt sh}^2,\quad{\cal B}_{47}={\tt ch}{\tt sh}^3,\quad
{\cal B}_{4,11}=-{\tt sh}^4\nonumber\\
&&{\cal B}_{67}=2{\tt ch}^2{\tt sh}^2,\qquad{\tt ch}\equiv\cosh{\lambda},\quad{\tt sh}\equiv\sinh{\lambda}.
\end{eqnarray}
In both the cases the correspondence between ${\cal H}$ and ${\cal R}$ is given exactly by the formula \eqref{hr} (with $H,R\rightarrow{\cal H},{\cal R}$).

Let us study the case $h_{\rm st}=i$ in more details. Introducing the operators
\begin{eqnarray}
c_+(k)=\frac{c_1(k)+\gamma c_2(k)}{\sqrt{1+|\gamma|^2}},\qquad
c_-(k)=\frac{\bar\gamma c_1(k)+c_2(k)}{\sqrt{1+|\gamma|^2}},\qquad\gamma\equiv\frac{1+{\rm e}^{ik}}{2i(1-\sin{k/2})},
\end{eqnarray}
and using the identities
\begin{equation}
|\gamma|^2=\frac{\cos^2{k/2}}{(1-\sin{k/2})^2},\qquad1+|\gamma|^2=\frac{2}{1-\sin{k/2}},\qquad1-|\gamma|^2=-\frac{2\sin{k/2}}{1-\sin{k/2}},
\end{equation}
one readily gets from \eqref{comc12} and \eqref{Hc}
\begin{equation}\label{compm}
\{c_{\pm}^{\dagger}(k),c_{\pm}(p)\}=\delta(k-p),\qquad\{c_+^{\dagger}(k),c_-(p)\}
=i\frac{1+{\rm e}^{-ik}}{2}\delta(k-p),
\end{equation}
and
\begin{eqnarray}\label{Hi}
&&\hat H(i)=\int_0^{2\pi}\frac{dk}{\sin{k/2}}\Big[i(c_+^{\dagger}(k)c_+(k)-
c_-^{\dagger}(k)c_-(k))\nonumber\\
&&-\frac{1}{2}\Big((1+{\rm e}^{ik})c_+^{\dagger}(k)c_-(k)+(1+{\rm e}^{-ik})c_-^{\dagger}(k)c_+(k)\Big)\Big].
\end{eqnarray}

According to \eqref{compm} and \eqref{Hi},
\begin{equation}\label{[Hcp]}
[\hat H(i),c^{\dagger}_{\pm}(k)]=\pm i\sin{\frac{k}{2}}c^{\dagger}_{\pm}(k),
\end{equation}
and
\begin{eqnarray}\label{[Hcm]}
&&[\hat H(i),c_+(k)]=\frac{1}{\sin{k/2}}\Big[-i\Big(1+\cos^2{\frac{k}{2}}\Big)c_+(k)+(1+{\rm e}^{ik})c_-(k)\Big],\nonumber\\
&&[\hat H(i),c_-(k)]=\frac{1}{\sin{k/2}}\Big[(1+{\rm e}^{-ik})c_+(k)+i\Big(1+\cos^2{\frac{k}{2}}\Big)c_-(k)\Big].
\end{eqnarray}

As it follows from \eqref{[Hcp]} and \eqref{[Hcm]}, the spectrum of $\hat H(i)$ should be purely imaginary. Hence, the spectrum of $i\hat H(i)$ should be purely real.

At the present time an interest to the non-Hermitian quantum spin chains (and especially with real spectrums) has steadily grown (see \cite{9,10,11,12} and referenced therein). The author believes, that the suggested model also may be interesting.

\end{document}